\documentclass[11pt,a4paper]{article}

\usepackage[margin=3cm]{geometry}
\usepackage{color}
\usepackage{amsmath,amssymb,amsthm,amsopn,bm,mathrsfs}
\usepackage[en-GB]{datetime2}
\DTMlangsetup[en-GB]{ord=omit}
\usepackage[comma]{natbib}
\usepackage{algorithm}
\usepackage{algorithmicx}
\usepackage[noend]{algpseudocode}
\usepackage{graphicx}
\usepackage[title]{appendix}
\usepackage[hidelinks]{hyperref} %% best to load hyperref as last package to avoid macro conflicts

\newcommand{\bb}{{\bm b}}

\newcommand{\be}{{\bm e}}
\newcommand{\bv}{{\bm v}}
\newcommand{\bbm}{{\bm m}}

\newcommand{\bbf}{{\bm f}}
\newcommand{\bg}{{\bm g}}

\newcommand{\bk}{{\bm k}}

\newcommand{\bt}{{\bm t}}
\newcommand{\rref}{{\rm ref}}
\newcommand{\cand}{{\rm cand}}
\newcommand{\candt}{{\rm ct}}

\newcommand{\pkg}[1]{{\tt #1}}
\newcommand{\proglang}[1]{{\sf #1}}
\newcommand{\code}[1]{{\tt #1}}
\newcommand{\Start}{\operatorname{Start}}
\newcommand{\End}{\operatorname{End}}
\newcommand{\Edges}{{\rm Edge}}
\newcommand{\prev}{{\rm prev}}

\title{Using iterated local alignment to aggregate trajectory data into a traffic flow map}

\author{Tarn Duong\footnote{Paris, France F-75000. Email:
{\tt tarn.duong@gmail.com}. ORCID: \href{https://orcid.org/0000-0002-1198-3482}{0000-0002-1198-3482}}}

\begin{document}

\maketitle

\begin{abstract}
\noindent Vehicle trajectories are a promising GNSS (Global Navigation Satellite System) data source to compute multi-scale traffic flow maps ranging from the city/regional level to the road level. The main obstacle is that trajectory data are prone to measurement noise. While this is negligible for city level, large-scale flow aggregation, it poses substantial difficulties for road level, small-scale aggregation. To overcome these difficulties, we introduce innovative local alignment algorithms, where we infer road segments to serve as local reference segments, and proceed to align nearby road segments to them. We deploy these algorithms in an iterative workflow to compute locally aligned flow maps. By applying this workflow to synthetic and empirical trajectories, we verify that our locally aligned flow maps provide high levels of accuracy and spatial resolution of flow aggregation at multiple scales for static and interactive maps.   
\\ \parskip \parskip

\end{abstract}

\section{Introduction}

One of the fundamental quantities in transport planning is a traffic flow map, i.e., a map of the traffic flow levels on the road segments in a road network \citep{ortuzar2011}. Whilst traffic flow maps are a rich source of information about vehicle mobility patterns, they are costly in terms of time and resources to compute for any reasonably sized road network. To alleviate this cost burden, most approaches restrict the spatial resolution of the flow map. One of the most common is to place sensors at fixed locations in the road network, whose results are then visualized as a traffic count map. Another common approach are trip intent/recall questionnaires to inform large-scale properties such as trajectory origins and destinations, whose results are then visualized as a desire line flow map/spider diagram \citep{tobler1987}.
Both of these are simplified flow maps, since the detailed mobility patterns outside of the sensor locations or origin/destination pairs are not known. These unknown  patterns can be inferred from other data sources, such as route assignment models \citep{evans1976}. While these model-assigned routes are highly detailed, the trade-off is that they are not guaranteed to correspond closely to empirical mobility patterns. 

\begin{figure}[!ht]
\includegraphics[width=\textwidth]{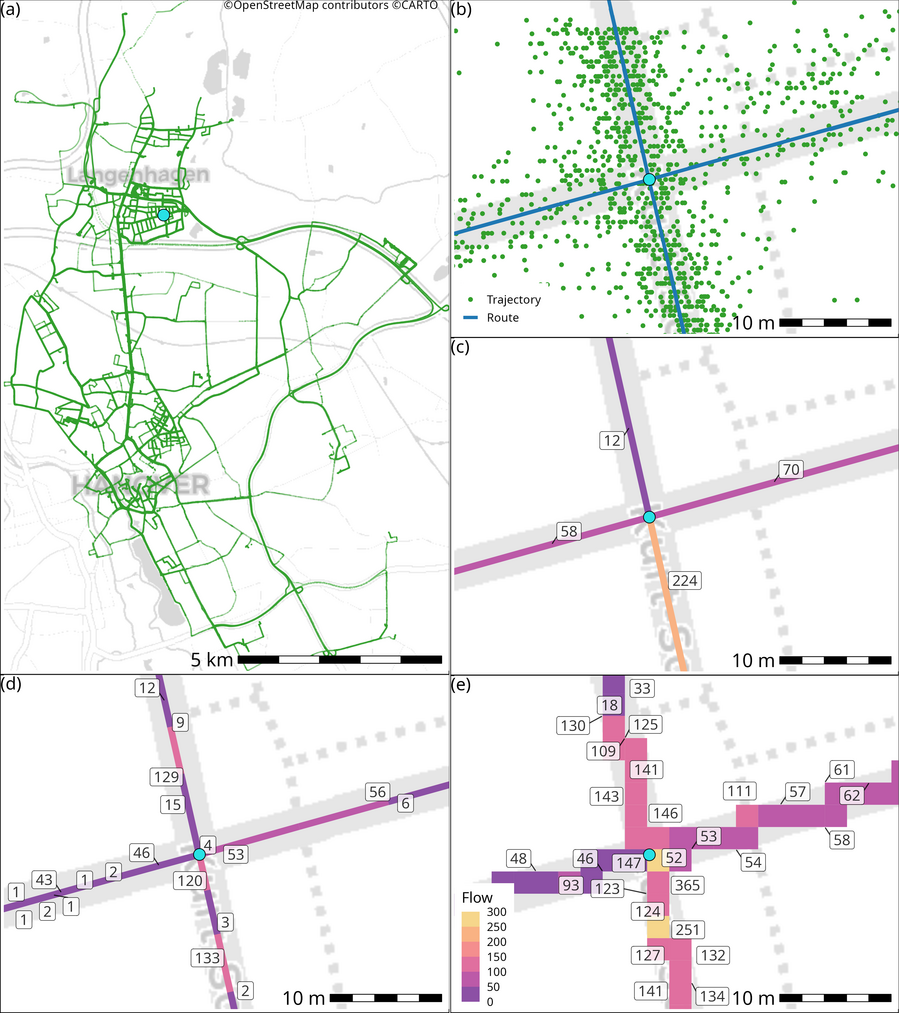} 
\caption{Trajectory data in Hannover, Germany. (a) Empirical trajectories (green circles), Loebensteinstraße (solid cyan circle). (b) Zoom of trajectories and map matched routes (blue lines). (c) Zoom of ideal flow aggregation. (d) Zoom of existing flow aggregation. (e) Zoom of 2\,m $\times$ 2\,m rasterised flow aggregation. Colour scale/label is traffic flow in road segments/grid cells. }
\label{fig:traj1}
\end{figure}

Trajectory data, derived from GNSS (Global Navigation Satellite Systems), combine the large-scale information of road sensor counts or questionnaires, with the small-scale details of model-assigned trajectories. Due to the prevalence of GNSS-enabled devices, such as vehicle navigation guides and mobile telephones, trajectory data can be acquired with low marginal cost whilst at the same time offering extensive spatial coverage and resolution of empirical mobility patterns \citep{herrera2010,andrienko2013}.
Our motivating example of trajectory data are the 1183 trajectories in EPSG 32632 (WGS 84/UTM zone 32N) projected coordinates, collected from a GNSS-enabled mobile phone, from December 2017 to March 2019 by a single driver in Hannover, Germany, with an overall average sampling rate about 1 point per second \citep{zourlidou2022}. In Figure~\ref{fig:traj1}(a), at the city level, the green circles appear to be aligned to the road network.

If we zoom in on these trajectory points around Loebensteinstraße (solid cyan circle), formerly known as Hindenburgstraße, is a residential street near the A2 autobahn, then we observe that in Figure~\ref{fig:traj1}(b) the points deviate from the road network. These noisy trajectories are unsuitable for direct aggregation into a flow map, so the usual first step is to apply a map matching procedure \citep{quddus2007,chao2020}. 
The map matched routes are the  blue lines in Figure~\ref{fig:traj1}(b). These follow the road network more closely. Map matching is a highly difficult problem and map matching with exact alignment to the underlying road network in a reasonable time frame is usually not feasible. The approach for fast computations taken by the Valhalla routing engine introduces some approximations which do not lead to exact alignment of empirical trajectories with the road network or to each other \citep{sghaier2019}. Nonetheless the misalignment is reduced to less than 5\,m, which is, for many purposes, including most notably navigation, indeed a suitable result.  

The ideal flow aggregation from the empirical trajectories is in Figure~\ref{fig:traj1}(c). The colour scale for the flow ranges from 0 (violet) to 150 (pink) to 300 (yellow). Applying the state-of-the-art traffic flow aggregation API \citep{morgan2021} to the inexactly aligned trajectories leads to an inaccurate flow map in Figure~\ref{fig:traj1}(d). 
The authors resolve this by rasterising the flow map to a raster grid: within each raster grid cell, the flow is the sum of the flows of all trajectories which intersect the grid cell. This rasterised flow map is displayed in Figure~\ref{fig:traj1}(e).  The 2\,m $\times$ 2\,m raster grid obscures the road network at this scale. The resolution of a rasterised aggregation flow map is fixed by the resolution of the raster grid. If we require a higher resolution, then we must recalculate the flow map on a raster grid with a higher resolution. 

Our goal is to produce a traffic flow map from empirical trajectories which is aligned to the road network and which can be utilised at any scale without recalculation, from the city level to the individual road level. In our proposed local alignment approach, we infer which road segments should serve as a local reference, and then proceed to align other nearby road segments to it. To accomplish this, we introduce a novel algorithm ``line blending'' where road segments, which are near each other but do not share overlapping sub-segments, are aligned to maximise their overlapping sub-segments. Line blending is a ``fuzzy'' aggregation method which calculates accurate flow aggregation of noisy trajectories. This is in contrast to the current state-of-the-art flow aggregation methods which are accurate only for noise-free trajectories. We combine these novel algorithms into an iterative workflow which outputs a locally aligned flow map. This flow map is composed of road segments rather than raster grid cells, so its resolution is not limited by the raster grid resolution. 

The outline of the paper is as follows. 
In Section~\ref{sec:align-local}, we construct an iterative workflow which computes a locally aligned flow map from noisy trajectories. In Section~\ref{sec:discussion}, we apply this workflow to synthetic trajectories from Los Angeles, USA and empirical trajectories from Hannover, Germany. We verify that, with a suitable tuning parameter choice, the locally aligned flow maps have high resolution and spatial accuracy. We then discuss some computational issues and end with some concluding remarks. 

\section{Local alignment of route segments for flow aggregation}
\label{sec:align-local}

We represent the road network by a graph $\mathcal{N} = \mathcal{N}(\mathcal{E}, \mathcal{V})$, where the $n_\mathcal{E}$ edges $\mathcal{E} = \{\be_1, \dots, \be_{n_\mathcal{E}}\}$  are the road segments and the $n_\mathcal{V}$ nodes/vertices $\mathcal{V} = \{\bv_1, \dots, \bv_{n_\mathcal{V}} \} $ indicate that different road segments are accessible to/from each other at this node point. We set  $\mathcal{N}$ to be the OpenStreetMap (OSM) road network. Each path in the network is composed of a connected sequence of edges, and we refer to this as a ``linestring'' geometry. We denote a single trajectory $G = \{\bg_1, \dots, \bg_{n_G}\} $ as a temporally ordered sequence of $n_G$ points $\bg_i, i= 1, \dots, n_G$. This is a ``multipoint'' geometry. 
The points of a trajectory $G$ are not necessarily coincident with the road network $\mathcal{N}$.

\subsection{Map matching}

We represent the output of a map matching algorithm $M$ from a trajectory $G$ as $M(G) = \{\bbm_1, \dots, \bbm_{n_M}\}$ which is a connected sequence of $n_M$ edges, i.e. a linestring geometry. 
In comparison to a trajectory $G$, all boundary points of the edges of a map matched route $M(G)$ are aligned to the road network.  
There is a large body of research on this difficult problem of map matching. 
We focus on the popular class of Hidden Markov Models (HMM) map matching algorithms. HMM methods iteratively build the map matched route by selecting the most likely next segment to connect to the current route using a probabilistic model. 
According to a review of map matching algorithms \citep{chao2020}, HMM is a state-transition method. The other three classes are similarity, candidate-evolving, and scoring methods. Further details of these alternatives are found in \cite{quddus2007,chao2020}. We leave this discussion here since our proposed methods are valid for any map matching algorithm, and concentrate on HMM map matching due to its accuracy and computational efficiency.
Even if we restrict ourselves to HMM algorithms on the OSM road network, there are many off-the-shelf map matching APIs available. We focus on the Valhalla routing engine (\url{https://valhalla.github.io/valhalla}), which includes its highly recommended map matching API \citep{saki2022}.
The inputs to \code{ST\_ROUTE} are the trajectory $G$ and the map matching API $M$. For readability, we defer the detailed description of the algorithm (and all subsequent algorithms) to the Appendix.

\subsection{Flow aggregation}

To the road network graph $\mathcal{N} = \mathcal{N}(\mathcal{E}, \mathcal{V})$, we add the traffic flows on each of the network edges to form a flow map $\mathcal{F} = \{ (f, \bbf) : \bbf \in \mathcal{N}, f > 0\}= \{(f_1,\bbf_1), \dots, (f_{n_\mathcal{F}}, \bbf_{n_\mathcal{F}}) \}$ composed of $n_\mathcal{F}$ flows, 
where each $\bbf_i$ is composed of edges in $\mathcal{E}$, with traffic flow $f_i$, $i=1, \dots, n_\mathcal{F}$. Our goal is to estimate the traffic flows $f_1, \dots, f_{n_\mathcal{F}}$ and the flow network graph $\mathcal{F}$. For exactly aligned trajectories, to compute the flow aggregation, we deploy the \code{overline} function in the \proglang{R} package \pkg{stplanr}, which we refer to as \code{ST\_OVERLINE\_PLANR} \citep{lovelace2018}. That is, starting with the map matched routes $\mathcal{M} = \{ M(G_1), \dots, M(G_n)\}$, the flow map is $\mathcal{F} = \code{ST\_OVERLINE\_PLANR}(\mathcal{M})$. 

To compute an accurate flow map from noisy trajectories relies on resolving the crucial problem of how to aggregate similar, but not exactly overlapping, road segments. Historically most solutions have focused on eliminating the discrepancies in the map matching so that flow aggregation then becomes a straightforward procedure, drawing on the many map matching algorithms available. Despite this research effort, map matching algorithms which yield error-free output are limited to special cases, and none have the same  availability and geographical coverage as the open source, global Valhalla map matching. Following the recommended best practice, we carry out some preprocessing to simplify the map matched routes to improve the flow aggregation. We highlight two existing methods, which we call node snapping and node splitting.

Node snapping is where the boundary points of the traffic flow linestrings are snapped to each other. Since the former are also nodes of the flow map, this gives the name to the algorithm. We focus on snapping these nodes since the linestring misalignments are in part caused by the existence of nodes which are close to each other but not exactly equal. Since we are searching for points which are close to together, then this is well-suited to  statistical clustering. 
There are many statistical clustering algorithms available, and we focus on hierarchical clustering. To resolve the computational bottleneck of hierarchical clustering, we approximate the 1-pass complete linkage clustering by a nested 2-pass clustering. We begin with an efficient single linkage clustering in the \proglang{R} package \pkg{fastcluster} \citep{mullner2013}. 
Since single linkage can result in chain-like clusters, we compute a subsequent complete linkage clustering to break these potential chains. In this nested approach, the complete linkage distance matrix is calculated only within each single linkage cluster, and so it is less likely to reach computational limits. This approximation is required since \pkg{fastcluster} does not implement an efficient complete linkage clustering.       
The inputs to  Algorithm~\ref{alg:st-snapnode} (\code{ST\_SNAPNODE}) are the flow linestrings $\mathcal{F}$ and the snap tolerance $\varepsilon_S$. The output are the node snapped linestrings $\mathcal{F}^*$. As its name suggests, all node points of $\mathcal{F}$ within $\varepsilon_S$ distance of each other are snapped together. 
In comparison to the unsnapped linestrings $\mathcal{F}$, there are fewer snapped linestrings $\mathcal{F}^*$ which have higher flow values, are more interconnected, and share more exactly overlapping sub-segments. 

Node snapping is focused near the boundary points of the flow linestrings, so it does not address misalignments far from the boundary points.
Moreover there remain some linestrings whose intersection is not correctly computed by \code{ST\_OVERLINE\_PLANR}. The solution to both of these problems is the explicit addition of the missing intersection nodes to these linestring, and then split these linestrings into simple linestring segments at these added nodes. This is known as node splitting. 
The inputs to Algorithm~\ref{alg:st-splitnode} (\code{ST\_SPLITNODE)} are the flow linestrings $\mathcal{F}$ and the node splitting type $S$. Our method deploys for $S=$ ``subdivision'' \code{to\_spatial\_subdivision} in the \pkg{sfnetworks} package \citep{sfnetworks}, and for $S=$ ``unary'' \code{geos\_unary\_union} in the \pkg{geos} package \citep{geos}. The first option tends to find fewer intersection nodes and is similar to the intersection computation performed by \code{ST\_OVERLINE\_PLANR}. The second option finds more missing intersection nodes, and leads to  more comprehensive flow aggregation.  

\begin{figure}[!ht]
\includegraphics[width=\textwidth]{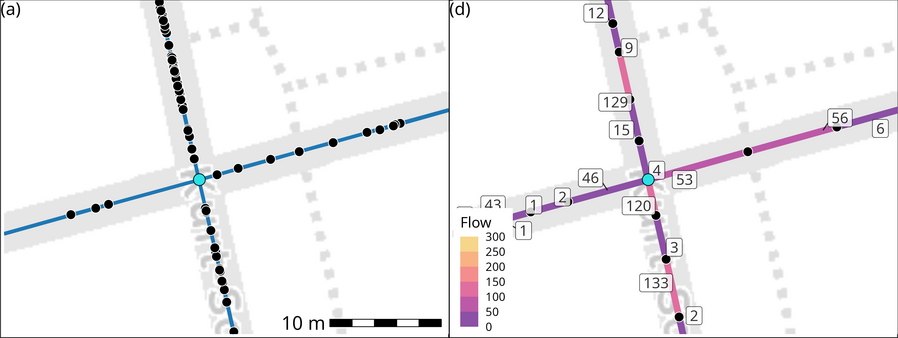} 
\caption{Unsplit/unsnapped and node split/node snapped routes. (a) Unsplit and unsnapped routes. (b) Node split and node snapped, with snap tolerance $\varepsilon_S=4$\,m and with flow aggregation. Colour scale/label is traffic flow in road segments. }
\label{fig:flow3}
\end{figure}

In Figure~\ref{fig:flow3}(a) are the map matched routes without any node splitting or node snapping. There are missing intersection nodes, and there are intersection nodes (solid black circles) which are close to each other. In Figure~\ref{fig:flow3}(b) is the flow aggregation after node splitting ($S=$ ``unary'') and then node snapping (with snap tolerance $\varepsilon_S=4$\,m). The result is that there are fewer flow segments with higher flow values. Due to the combined action of \code{ST\_SPLITNODE} and \code{ST\_SNAPNODE}, Figure~\ref{fig:flow3}(b) is closer to the ideal flow map topology than Figure~\ref{fig:flow3}(a). 

Whilst the preprocessing does improve the flow aggregation, there remain some discrepancies, since Figure~\ref{fig:flow3}(b) is not quite the ideal flow map. To tackle directly to the problem of eliminating the discrepancies during flow aggregation, heuristic solutions have most likely been investigated, though few have been published so there is scarce literature on this problem. 
Among the few published solutions include edge bundling \citep{zhou2013} and rasterisation \citep{wood2010,morgan2021}. Edge bundling consists of clustering trajectory linestrings and replacing all cluster members with a single representative linestring. These (and subsequent) authors conclude that it performs poorly when applied to noisy trajectories at the road level, and remains mostly suited to large-scale aggregations, such as desire line maps \citep{tobler1987}. We do not consider edge bundling further since it is not suitable for our purposes. Rasterisation relies on converting the flow map into a raster matrix, and aggregating the flows within the same raster pixel neighbourhood. Whilst rasterisation is able to improve flow aggregation, it depends highly on the raster pixel neighbourhood size, and leads to a loss of spatial resolution. 

We propose an alternative aggregation which solves both of these problems. We achieve this goal by local internal alignment between the map matched routes. By internal alignment, we mean that we align the routes with each other, rather than to the external road network graph. 
By local alignment, we mean that we align segments of the routes, rather than the complete routes.  Our proposal is  developed within the \proglang{R} statistical analysis environment due to its integrated access to advanced statistical and geospatial analysis methods.

\subsection{Line blending to align similar linestrings to local reference linestring}

So far we have focused on improving the alignment of linestrings by resolving inconsistencies at their intersections. We now focus on aligning linestrings more generally.
For this, we require a comparison relationship to establish an order of alignment of nearby linestrings. For two flow linestrings $(f_1,\bbf_1)$ and $(f_2,\bbf_2)$ from a flow map $\mathcal{F}$, we define that $\bbf_2$ is a candidate to be aligned to the reference $\bbf_1$ at threshold $\varepsilon \geq 0$ if 
\begin{equation} 
\bbf_2 \subseteq \code{ST\_BUFFER}(\bbf_1, \varepsilon), \ f_2 \leq f_1. 
\label{eq:lineblend}
\end{equation}
The condition on the flow values $f_2 \leq f_1$ means that we place a higher priority on the reference linestring $\bbf_1$ since its flow $f_1$ is larger than the flow $f_2$ on the candidate linestring $\bbf_2$.  
The buffer zone has flat edges, e.g. \code{ST\_BUFFER(endCapStyle="FLAT")} in the \proglang{R} package \pkg{sf} \citep{sf}, so the buffer zone ends at the boundary points of $\bbf_1$. This avoids erroneously considering neighbouring segments of $\bbf_1$, which are connected to $\bbf_1$ as a part of a longer sequence, to be candidates to be aligned to $\bbf_1$. 

Let $\bbf_\rref$ be a reference linestring from $\mathcal{F}$. The set of $n_{\mathcal{F}_\cand}$ linestrings from $\mathcal{F}\backslash \{\bbf_\rref\}$ which satisfy Equation~\eqref{eq:lineblend} are the candidate linestrings $\mathcal{F}_\cand = \{ \bbf_{\cand,1}, \dots, \bbf_{\cand,n_{\mathcal{F}_\cand}} \}$, with the convention that if $n_{\mathcal{F}_\cand}=0$ then $\mathcal{F}_\cand$ is the empty set. 
We call our approach ``line blending'', since we blend the candidate linestrings onto the reference linestring. This line blending resolves the crucial problem of how to aggregate similar, but not exactly overlapping, road segments. This ``fuzzy'' aggregation allows us to compute accurate flow aggregation from noisy trajectories.  

The inputs in Algorithm~\ref{alg:st-lineblend} (\code{ST\_LINEBLEND}) are the reference linestring $\bbf_\rref$, the candidate linestrings $\mathcal{F}_\cand$, and the blend tolerance $\varepsilon$. The output are the modified reference linestring $\bbf^*_\rref$ and projected candidate flow linestrings  $\mathcal{F}^*_\cand$, all with added interior points for accurate calculation of exactly equal linestring segments. If there are many candidate linestrings, then this may lead to many sub-segments in the projected linestrings, each with their own flow values. So we assign the rounded value of the weighted mean flow, weighted by the sub-segment length, to all sub-segments. 

\begin{figure}[!ht]
\includegraphics[width=\textwidth]{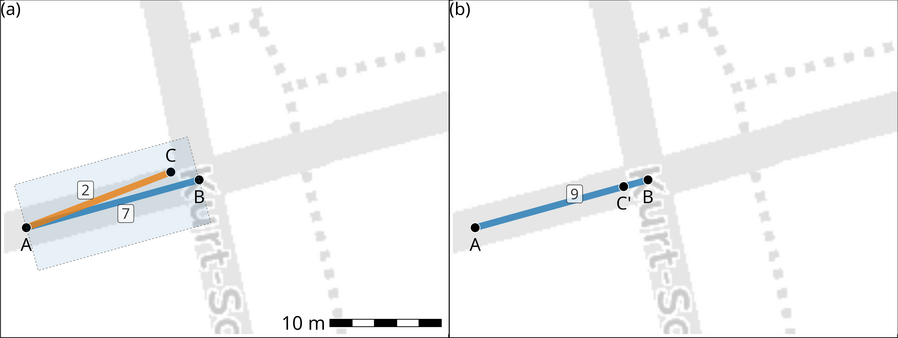}
\caption{Line blending leads to correct flow aggregation. (a) Before line blending and flow aggregation. Reference linestring $(7,ABC)$ in blue, candidate linestring $(2,AD)$ in orange. (b) After blending candidate into reference linestring, with blend tolerance $\varepsilon=4$\,m, and flow aggregation. Blended linestring is $(9,AC'B)$.  } 
\label{fig:flow4}
\end{figure}

In Figure~\ref{fig:flow4}(a), we illustrate this line blending with the flow linestrings $(7,AB)$ in blue and $(2,AC)$ in orange. Since $AD \subset \code{ST\_BUFFER}(AB, 4\,\mathrm{m})$ (pale blue rectangle), and $2 < 7$, then $AC$ is the reference $\bbf_\rref$ and $AD$ is the candidate linestring $\bbf_\cand$, according to Equation~\eqref{eq:lineblend}.
Line blending involves projecting $AD$ to $AC$, so $C$ is projected to $C'$ (which lies exactly on the reference linestring). We also add $C'$ to the reference linestring. The reference linestring becomes $\bbf^*_\rref= (7, AC'B)$, and the candidate $\bbf^*_\cand= (2, AC')$. Now $\bbf^*_\rref, \bbf^*_\cand$ share the sub-segment $AC'$ exactly, and \code{ST\_OVERLINE\_PLANR} gives the aggregated flows as $(9,AC')$ and $(7,C'B)$. We take the rounded value of the weighted mean of these two flow linestrings. The result is a single linestring $(9,AC'B)$, as shown in Figure~\ref{fig:flow4}(b). Line blending is required to aggregate correctly  the linestrings $(7,AC)$ and $(2,AD)$ since they do not share a common sub-segment.

More complex situations arise when other linestrings touch the candidate linestring, but are not candidates themselves for blending. 
Since line blending projects the candidates to the reference linestring, then we have to also project these other linestrings to avoid leaving a gap in the updated flow map.
The inputs to Algorithm~\ref{alg:st-snap-cand-touch} (\code{ST\_SNAP\_CAND\_TOUCH}) are the reference linestring $\bbf_\rref$, the candidate-touching linestrings $\mathcal{F}_\candt$, and the snap tolerance $\varepsilon_S$. The output are the snapped candidate-touching flow linestrings  $\mathcal{F}^*_\candt$, which maintain connectivity between $\mathcal{F}^*_\candt$ and $\bbf_\rref$.

\begin{figure}[!ht]
\includegraphics[width=\textwidth]{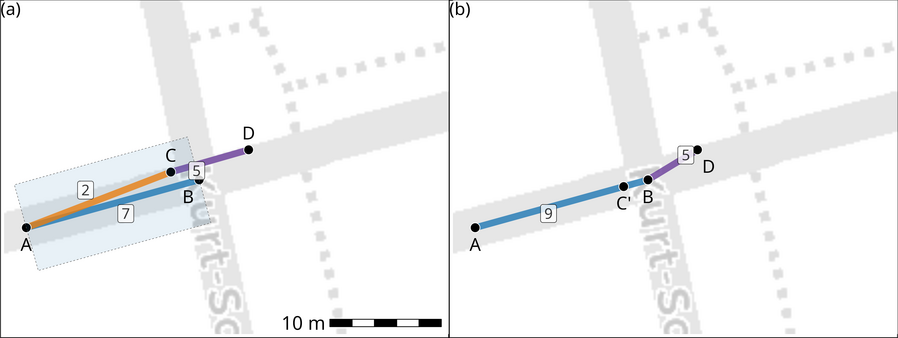} 
\caption{Snap candidate-touching linestrings to reference linestring after line blending. (a) Before line blending and snapping. Reference linestring $(7,AC)$ in blue, candidate linestring $(2, AC)$ orange, and candidate touching linestring $(5,CD)$ purple. (b) After line blending (blend tolerance $\varepsilon=4$\,m) and snapping (snap tolerance $\varepsilon_S=4$\,m). Reference linestring becomes $(9,AC'B)$, and candidate-touching linestring $(5,BD)$. }
\label{fig:flow5}
\end{figure}

Figure~\ref{fig:flow5}(a) is an illustration of \code{ST\_SNAP\_CAND\_TOUCH} with the reference $(7,AB)$ (blue), the candidate linestring $(2,AD)$ (orange), and the blending buffer zone with blend tolerance $\varepsilon=4$\,m (pale blue rectangle). The candidate-touching linestring $\bbf_\candt=(5,CD)$ (purple) touches the candidate $AC$ at $C$. 
The orange line is entirely within the pale blue buffer zone, whereas the purple line extends outside of the buffer zone and so is not a candidate for blending to the reference linestring. 
If we apply \code{ST\_LINEBLEND} to blend the candidate linestring, then $C$ is projected to $C'$ to lie on the reference linestring, i.e., $\bbf_\cand^*=(2, AC')$. 
If we apply \code{ST\_SNAP\_CAND\_TOUCH} with snap tolerance $\varepsilon_S=4$\,m to the candidate-touching linestring, then the intersection point $C$ is projected to the boundary point $B$, i.e., $\bbf^*_\candt=(5, BD)$.    
Thus line blending, candidate-touching snapping, and flow aggregation yields that $\bbf^*_\candt=(5, BD)$ remains connected to $\bbf_\rref^* = (9, AC'B)$. 

\begin{figure}[!ht]
\includegraphics[width=\textwidth]{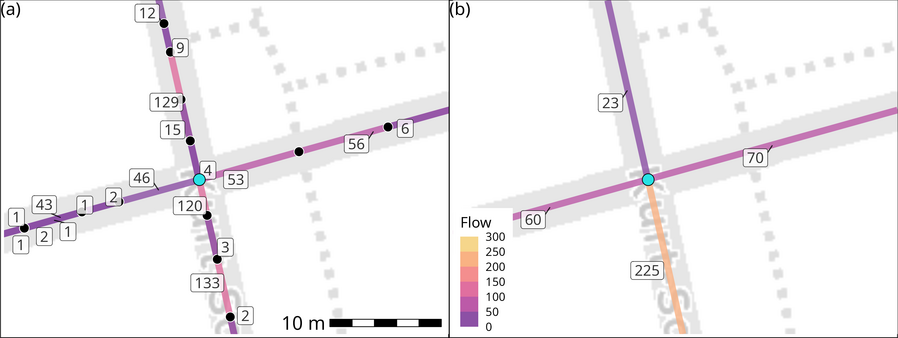} 
\caption{Hannover routes line blending. (a) Flow segments without line blending. (b) Flow segments with line blending. Colour scale/label is traffic flow. }
\label{fig:flow6}
\end{figure}

To illustrate line blending for the  Hannover map matched routes,  Figure~\ref{fig:flow6}(a) is the flow segments without any line blending, and  Figure~\ref{fig:flow6}(b) is the flow segments after line blending. Due to the action of \code{ST\_OVERLINE\_LINEBLEND}, each road segment in the line blended flow map Figure~\ref{fig:flow6}(b) corresponds unambiguously to a road segment in the OSM road network, and thus it approaches the ideal flow map. This is an improvement over the multiple road segments in the unblended flow map in Figure~\ref{fig:flow6}(a), and the pixelated road segments in the rasterised flow map in Figure~\ref{fig:traj1}(e). 

\subsection{Workflow}

Line blending for accurate flow aggregation at a single intersection in a road network is demonstrated in Figure~\ref{fig:flow6}. To extend line blending to cover the entire road network, we require two further generalisations. First, we generalise the reference linestring to be a sequence of $k$ connected edges, each with their own flow. If we treat this edge sequence momentarily as a single linestring, with flow equal to the weighted mean flow, weighted by the length of the $k$ edges, 
then we can apply Equation~\eqref{eq:lineblend} to search for potential candidate linestrings. Due to computational limitations, we retain that candidates be single edges with single flow values. The generalisation to $k$-edge reference linestrings allows us to blend candidate linestrings which exceed the buffer zones of 1-edge reference linestrings. 

Second, we determine the priority for line blending for a set of reference linestrings. We require that a reference linestring cannot be a candidate linestring to another reference linestring, and a candidate linestring is a candidate for one reference linestring only. These ensure that if two linestrings 
$\bbf_2 \subseteq \code{ST\_BUFFER}(\bbf_1, \varepsilon)$ and $\bbf_1 \subseteq \code{ST\_BUFFER}(\bbf_2, \varepsilon)$, then we select only one of them. The flow condition $f_2 \leq f_1$ usually distinguishes the reference linestring, except when $f_1=f_2$. In this case, then we designate the longer linestring to be the reference. 

The inputs to Algorithm~\ref{alg:st-lineblend-priority} (\code{ST\_LINEBLEND\_PRIORITY}) are the flow linestrings $\mathcal{F}$, the number of connected edges for the reference linestrings $k$, and the blend tolerance $\varepsilon$. The output is a set of non-overlapping reference linestrings $\mathcal{F}_\rref$, a set of unique candidate linestrings $\mathcal{F}_\cand$, and a set of (potentially repeated) candidate-touching $\mathcal{F}_\candt$ linestrings. 
By combining Algorithms~\ref{alg:st-snapnode}--\ref{alg:st-lineblend-priority}, we obtain an algorithm for local alignment of flow segments, in Algorithm~\ref{alg:st-overline-lineblend} (\code{ST\_OVERLINE\_LINEBLEND}). The inputs to \code{ST\_OVERLINE\_LINEBLEND} are the flow linestrings $\mathcal{F}$, the blend tolerance $\varepsilon$, and the snap tolerance $\varepsilon_S$. The output are the $\mathcal{F}^*$ aggregated aligned flow linestrings. 
  
To compute a locally aligned flow map, we combine Algorithms~\ref{alg:st-route}--\ref{alg:st-overline-lineblend} into an initial and an iterative phase. The inputs to Algorithm~\ref{alg:st-overline-initial} (\code{ST\_OVERLINE\_INITIAL}) are the map matched routes $\mathcal{M}$ and the split node type $S$. The output is an initial aligned flow map. 
With an initial flow map on hand, we proceed iteratively towards a locally aligned flow map. 
The inputs to this iterative step Algorithm~\ref{alg:st-overline} (\code{ST\_OVERLINE}) are the initial flow map $\mathcal{F}$, the split node type $S$, the blend tolerance $\varepsilon$, the snap tolerance $\varepsilon_S$, the number of connected edges $\bk$, and the maximum number of iterations $j_{\max}$. After a suitable number of iterations, the output is a locally aligned flow map.  

\begin{figure}[!ht]
\includegraphics[width=\textwidth]{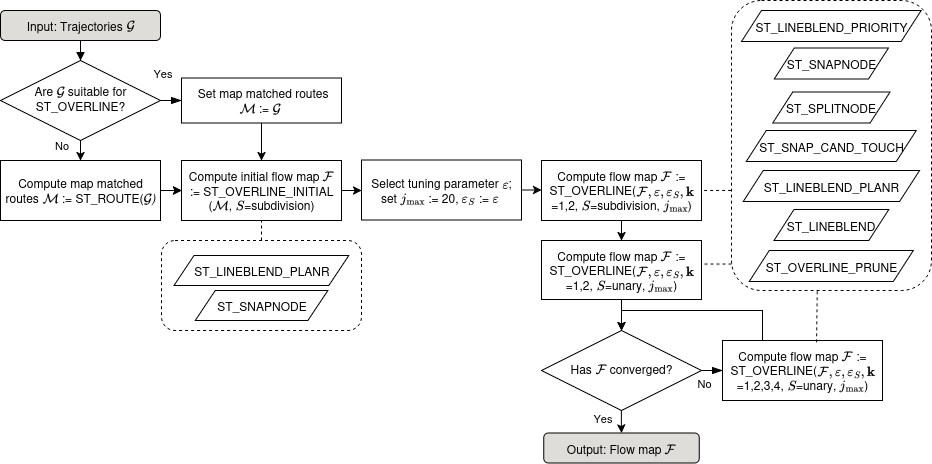} 
\caption{Workflow for locally aligned flow map. Input are trajectories $\mathcal{G}$, output is locally aligned flow map $\mathcal{F}$. Blend tolerance $\varepsilon$ is tuning parameter. } 
\label{fig:flowchart}
\end{figure}

Our proposed workflow from an input of trajectory data to a locally aligned flow map is illustrated in Figure~\ref{fig:flowchart}. We begin with the trajectories $\mathcal{G}$.
If the measurement errors in these trajectories are small enough to be input directly into \code{ST\_OVERLINE}, then we set $\mathcal{M} = \mathcal{G}$. Otherwise we carry out map matching with \code{ST\_ROUTE}. For the flow map aggregation, we begin with an iteration of \code{ST\_OVERLINE\_INIITAL} to create an initial flow map. In Figure~\ref{fig:flowchart}, the dotted rectangle surrounding \code{ST\_LINEBLEND\_PLANR} and \code{ST\_SNAPNODE} indicates that these algorithms are inputs to \code{ST\_OVERLINE\_INIITAL}. For the subsequent steps, we set the maximum number of iterations $j_{\max}=20$, and the snap tolerance $\varepsilon_S = \varepsilon$. 
We call \code{ST\_OVERLINE}  with \#edges $\bk=1,2$, and node split type $S=$ ``subdivision''.  
At this early stage, the search for connected $k$-edges with $k>2$ can be computationally intensive, and ``subdivision'' node splitting provides more stable line blending priority. The dotted rectangle  indicates the algorithms which are inputs to \code{ST\_OVERLINE}. We follow with a call to \code{ST\_OVERLINE} with ``unary'' node splitting. The ``unary'' node splitting is usually applied after ``subdivision'' node splitting since the former can now add any missing intersection nodes without adversely affecting the line blending priority. 
After these two steps, the flow map has begun to stabilise, and 3-, 4-edge searches become computationally feasible with the lower number of flow segments. We iterate \code{ST\_OVERLINE} with $\bk=1,2,3,4, S=$ ``unary'', until the flow map $\mathcal{F}$ converges. 
In this workflow, the only unspecified tuning parameter is the blend tolerance $\varepsilon$, whose properties we investigate in the next section. 

\section{Results and discussion}
\label{sec:discussion}

\subsection{Synthetic trajectories}

\begin{figure}[!ht]
\includegraphics[width=\textwidth]{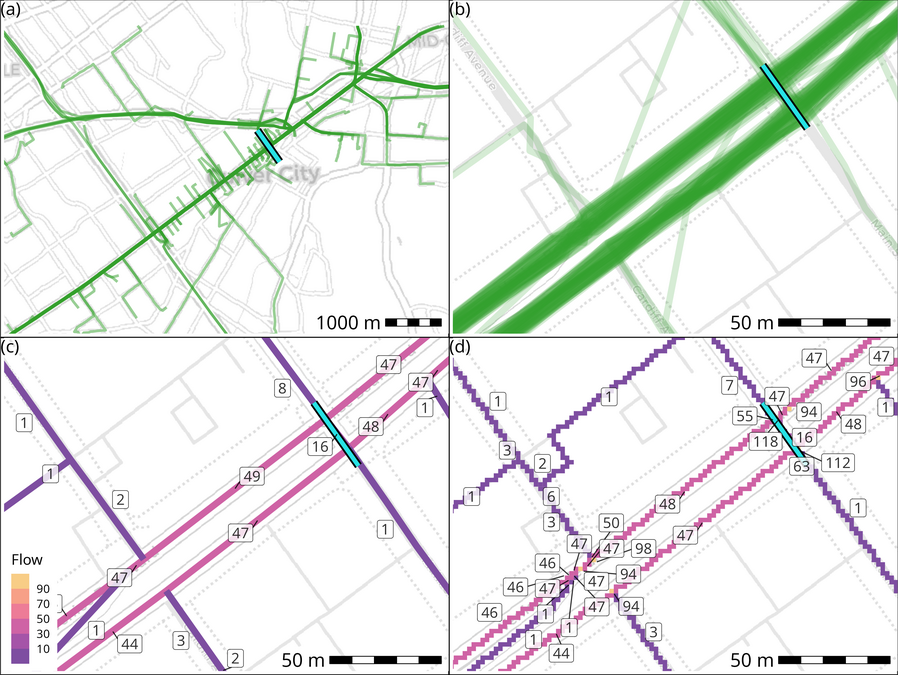}
\caption{Synthetic Culver City trajectories $(n=100)$. (a) Noisy trajectories (green lines). (b) Zoom of noisy trajectories near intersection of Main St and Venice Bd (solid cyan line). (c) Reference noise-free flow map. (d) Rasterised flow map with $\varepsilon_R=2$\,m. Colour scale/label is traffic flow. }
\label{fig:la-traj1}
\end{figure}

Despite the ubiquity of GNSS-enabled devices, there is a lack of gold standard empirical trajectories and flow maps, due to privacy concerns and other reasons. Our solution is to rely on noise-free synthetic trajectories which are exactly aligned to a theoretical road network \citep{anastasiou2022a}. These authors generate synthetic trajectories from an origin-destination matrix, and  make available a set of 1.5 million synthetic trajectories for the metropolitan area of Los Angeles, USA \citep{anastasiou2022b}.   
For illustrative purposes, we focus on a sample of $n=100$ trajectories, in EPSG 6339 (NAD83 (2011)/UTM zone 10N) coordinates, in Culver City (located about 16 km from downtown Los Angeles) which pass near the intersection of Main Street and Venice Boulevard, as indicated by the solid cyan line in Figure~\ref{fig:la-traj1}. 
We add a random uniform noise of $\pm5$\,m to the noise-free trajectories to obtain the noisy trajectories, displayed as the green lines in Figures~\ref{fig:la-traj1}(a--b). This noise mimics the typical measurement error of 5\,m observed in empirical trajectories from GNSS-enabled smartphones \citep{vandiggelen2015}. The reference noise-free flow map is in Figure~\ref{fig:la-traj1}(c), and the rasterised flow map in Figure~\ref{fig:la-traj1}(d). \cite{morgan2021} do not recommend a value for tuning the rasterisation grid resolution, so we select $\varepsilon_R=2$\,m subjectively so that the raster grid cells do not exceed the road widths. 

\begin{figure}[!ht]
\includegraphics[width=\textwidth]{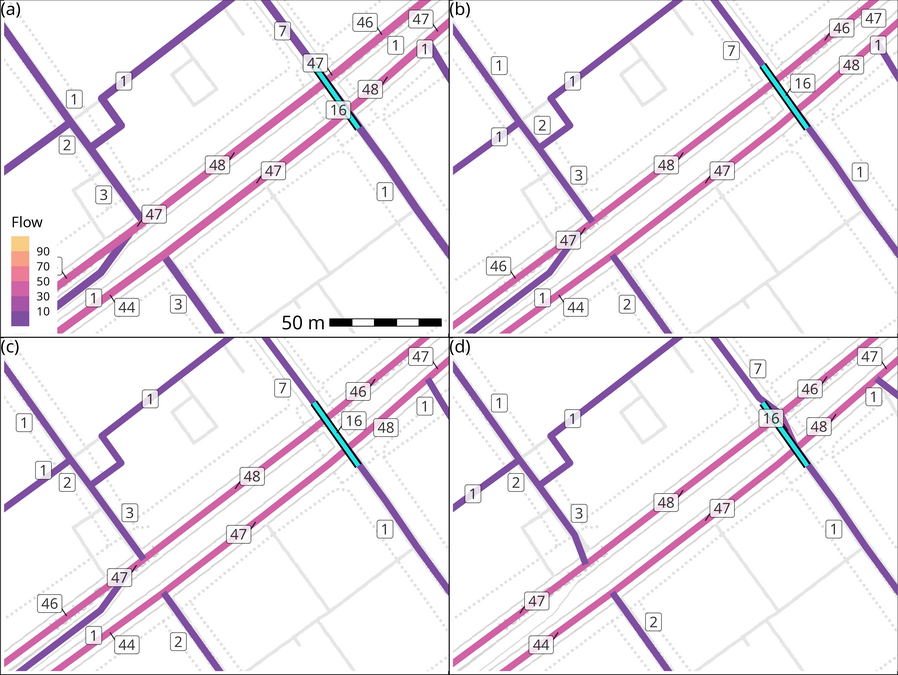}
\caption{Effect of blend tolerance $\varepsilon$ on synthetic Culver City flow maps. (a) Unblended flow map with $\varepsilon=0$\,m. (b) Flow map with $\varepsilon=1$\,m. (c) Flow map with $\varepsilon=5$\,m. (d) Flow map with $\varepsilon=10$\,m. Colour scale/label is traffic flow. }
\label{fig:la-flow1}
\end{figure}

For our locally aligned flow maps, the free tuning parameter is the blend tolerance parameter $\varepsilon$. So we investigate the effect of $\varepsilon$ on the accuracy of flow aggregation. For the blend tolerance $\varepsilon = 0$\,m, this means that no line blending is performed, so we compute an unblended flow map in Figure~\ref{fig:la-flow1}(a) which comprises many small road segments with low flow values. 
For the blend tolerance $\varepsilon = 1,5$\,m, the flow map in Figure~\ref{fig:la-flow1}(b--c), the flow maps are similar. Though we observe that $\varepsilon = 1$\,m leads to an unblended road segment with flow 1 (in the upper north west corner)m, since 1\,m is smaller than the remaining trajectory discrepancies.  
For the blend tolerance $\varepsilon = 10$\,m, near the intersection of Venice Bd and Main St (solid cyan line), the flow segment (with flow 16) is poorly aligned to the road network, unlike the more accurate alignment in Figures~\ref{fig:la-flow1}(b--c). 

We observe the usual sensitivity trade-off in the evolution of the values of the blend tolerance parameter. If this value is small (e.g. 0 or 1\,m in Figures~\ref{fig:la-flow1}(a--b)), then the action of line blending is limited to a small radius. The input trajectories are insufficiently processed, so the output flow map contains too many road segments. Whilst each discrepancy to the true road network is small (low bias), whenever the input trajectories are perturbed slightly, this would lead to a different flow map (high variability). If the blend tolerance is large (e.g. 10\,m in Figure~\ref{fig:la-flow1}(d)), the action of line blending is extended to a large radius. When the blend tolerance is larger than the distance between two distinct neighbouring road segments, then input trajectories which belong to different segments on the true road network can be erroneously blended together, and the output flow map contains an estimated road segment with a large discrepancy from any true road segment. Whilst the discrepancy is large (high bias), whenever the input trajectories are perturbed slightly, this would lead to a similar flow map (low variability). If the blend tolerance is between these two extremes (e.g. 5\,m in Figure~\ref{fig:la-flow1}(c)), then we achieve a trade-off between bias and variability. Therefore we set the blend tolerance to be sufficiently large so that it simplifies the flow map topology which approaches the true road network, and also not be too large so that it oversimplifies the flow map topology which deviates from the true road network.

To complement the above heuristic comparisons, we define a quantitative discrepancy between a flow map $\mathcal{F}$ and the reference flow map $\mathcal{F}_\rref$. First, we compute the transect points for each linestring $\bbf$ from $\mathcal{F}$: if $\operatorname{len}(\bbf) > \varepsilon_T$\,m, then a point is placed at every $\varepsilon_T$\,m of $\bbf$; or if $\operatorname{len}(\bbf) \leq \varepsilon_T$\,m, then a single point is placed at the centroid of $\bbf$. So $\mathcal{T} = \{\bt_{1}, \dots, \bt_{n_\mathcal{T}} \}$ is a set of $n_\mathcal{T}$ transect points. For each $\bt_i, i=1, \dots,n_\mathcal{T}$, we find the closest linestring in  $\mathcal{F}_\rref$, that is,  
$\bbf_{\rref,i}^* =  \operatorname{argmin}_{\bbf_\rref \in \mathcal{F}_\rref} \mathrm{dist}(\bt_i, \bbf_\rref)$. 
The flow errors are $\operatorname{Err}_i = |f_i - f_{\rref,i} ^*| $ for the transect points on $\bt_i \in \mathcal{T} \subseteq \mathcal{F}$. We reverse the order of the comparison to obtain the flow errors $\operatorname{Err}_j = |f_{\rref,j} - f_{j}^* | $ for the transect points on $\bt_j \in \mathcal{T}_\rref \subseteq  \mathcal{F}_\rref$, and concatenate these two sets of flow errors. 

\begin{figure}[!ht]
\includegraphics[width=\textwidth]{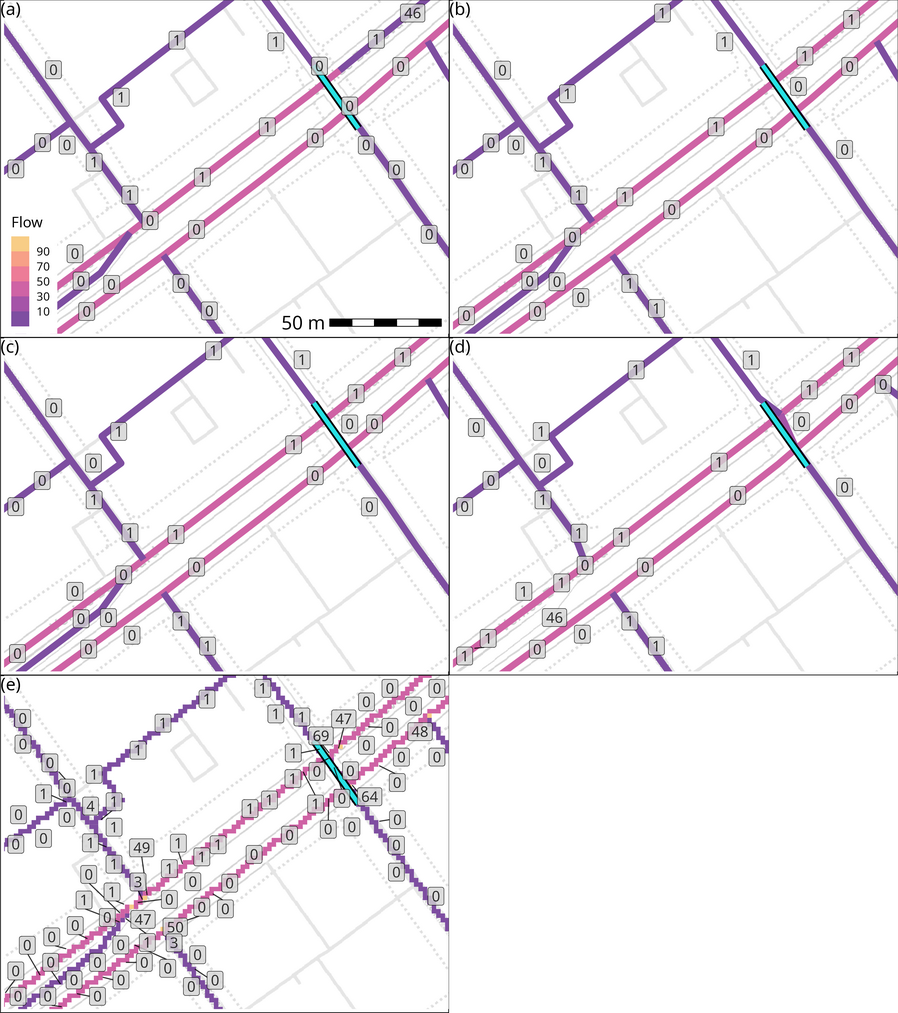}
\caption{Flow errors for synthetic Culver City flow maps. (a) Unblended flow map with $\varepsilon=0$\,m. (b) Flow map with $\varepsilon=1$\,m. (c) Flow map with $\varepsilon=5$\,m. (d) Flow map with $\varepsilon=10$\,m. (e) Rasterised flow map with $\varepsilon_R=2$\,m. Colour scale is traffic flow, label is traffic flow error. }
\label{fig:la-flow2}
\end{figure} 

In Figure~\ref{fig:la-flow2}, we display the transect points with $\varepsilon_T=50$\,m, as the grey circles, superposed over the flow map $\mathcal{F}$. Figure~\ref{fig:la-flow2}(a) is the unblended flow map, Figures~\ref{fig:la-flow2}(b--d) are the flow errors for the locally aligned flow maps with blend tolerance $\varepsilon=1,5,10$\,m, and Figure~\ref{fig:la-flow2}(e) is the rasterised flow map. The rasterised flow map lead to many high error values. Whereas for the locally aligned flow maps, the flow errors are low, especially for $\varepsilon=1, 5$\,m flow maps.

We repeat the flow map error calculations for $N=10$ re-samples of $n=100$ Culver City trajectories. The mean and SD of the flow errors for the unblended flow map is $0.386 \pm 0.048$ ($\varepsilon=0$\,m), for the locally aligned flow maps $0.351 \pm 0.055$ ($\varepsilon=1$\,m), $0.350 \pm 0.056$  ($\varepsilon=5$\,m), $0.374 \pm 0.061$ ($\varepsilon=10$\,m), and for the rasterised flow map $0.438 \pm 0.035$ ($\varepsilon_R=2$\,m).
Setting $\varepsilon=0$\,m leads to under-blending since $\varepsilon$ is smaller than the misalignments in the map matched routes, and setting $\varepsilon=10$\,m leads to over-blending since $\varepsilon$ is larger than the distance between separate road segments. The $\varepsilon_R=2$\,m rasterised flow map is similarly under-blended like the $\varepsilon=0$\,m flow map. So we tentatively recommend a value between 1 and 5\,m for the blend tolerance parameter $\varepsilon$ since it leads to optimal line blending. This optimal value is a trade-off between the magnitude of the misalignments in the empirical trajectories and the road density of the underlying road network.

\subsection{Empirical trajectories}

\begin{figure}[!htp]
\includegraphics[width=0.97\textwidth]{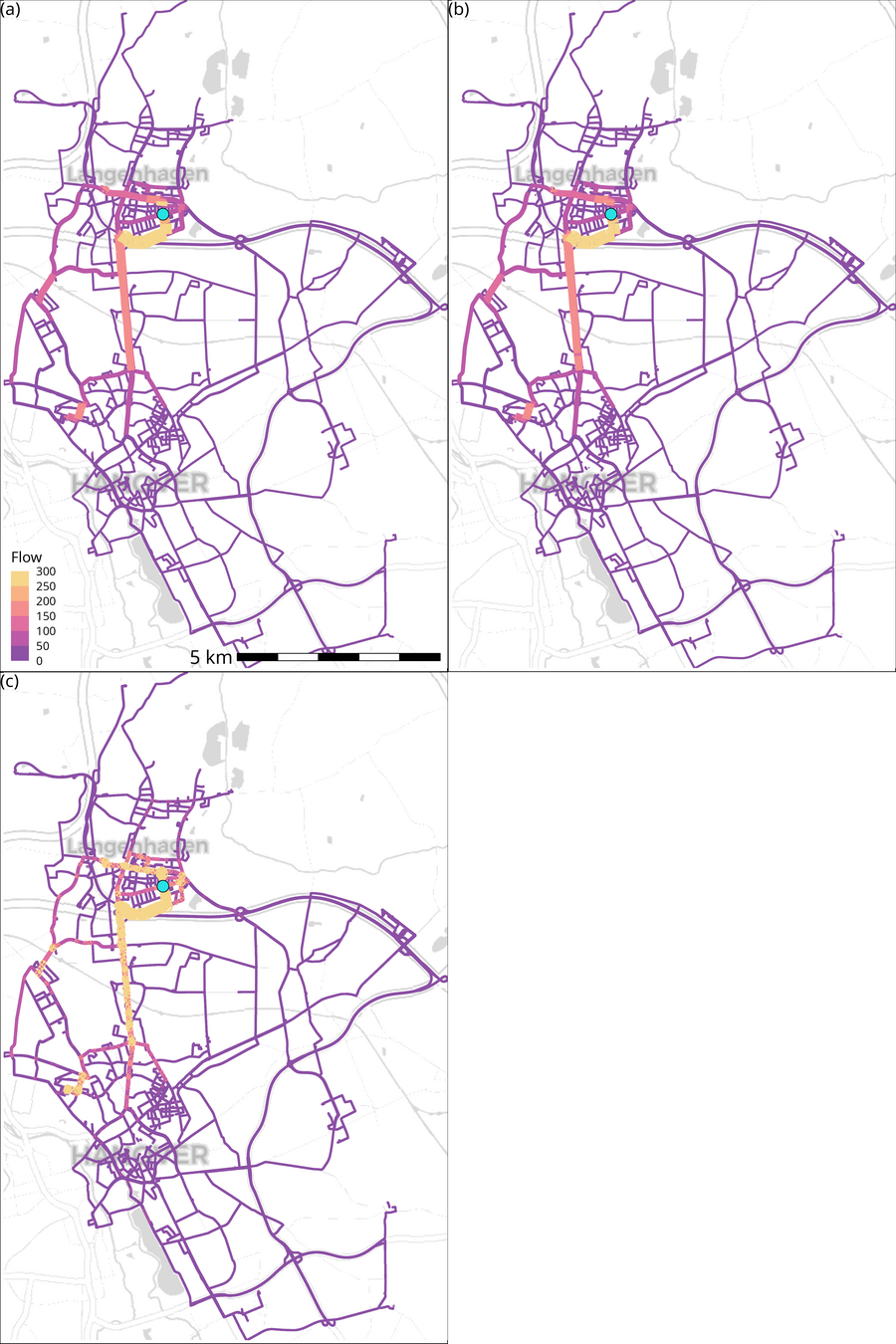} 
\caption{Hannover flow maps at city level. (a) Locally aligned flow map. (b) Unblended flow map.  (c) Rasterised flow map. Solid cyan circle is Loebensteinstraße. Colour scale is traffic flow. }
\label{fig:flow8}
\end{figure}

We return to the empirical Hannover trajectories. From the 1183 trajectories, we keep 1181 trajectories with length greater than $0$\,m, and follow the workflow in Figure~\ref{fig:flowchart}. We map match these 1181 trajectories with \code{ST\_ROUTE}, and we continue with 1176 of them which have a sufficiently high quality match, where the  Hausdorff distance $\operatorname{dist}_{\operatorname{Haus}}(R(G),G) < 100$\,m, and the ratio $\operatorname{len}(R(G))/\operatorname{len}(G) < 1.1$, for a trajectory $G$ and a map matched route $R(G)$. That is, we discard 5 trajectories or a 0.42\% failure rate for the Valhalla map matching, which is consistent with previous studies \citep{saki2022}.
After some trial and error, we set the blend tolerance of $\varepsilon=4$\,m, which is consistent with the tentative recommendation of $\varepsilon$ between 1 and 5\,m from the synthetic trajectories computations above. 
The workflow leads to a locally aligned flow map displayed in Figure~\ref{fig:flow8}(a). The corresponding unblended flow map is Figure~\ref{fig:flow8}(b) and the rasterised flow map is Figure~\ref{fig:flow8}(c). At the city level, all these flow maps visualise well the traffic flows.

\begin{figure}[!ht]
\includegraphics[width=\textwidth]{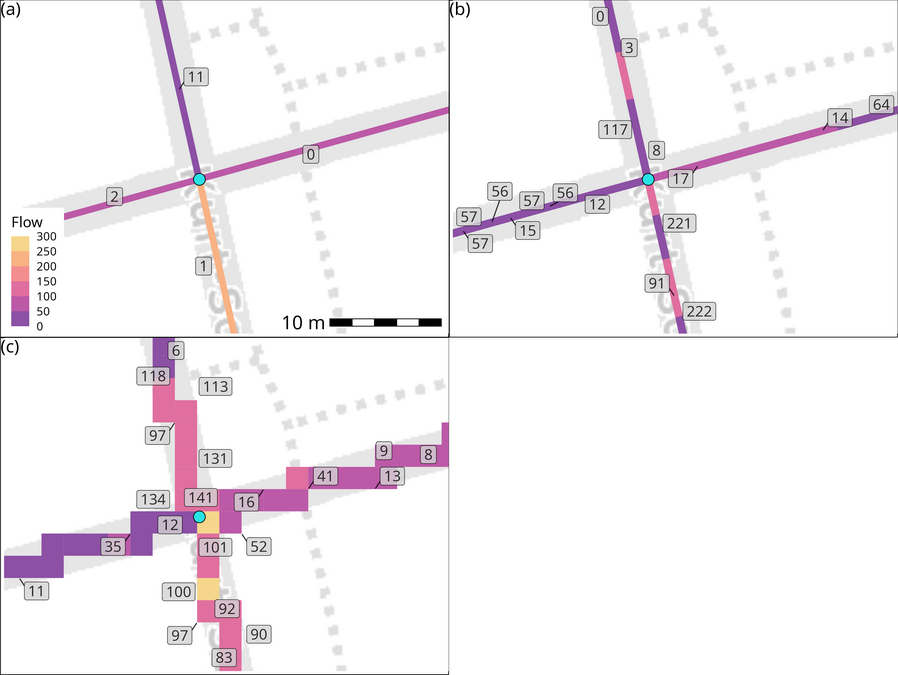} 
\caption{Hannover flow maps at road segment level. (a) Locally aligned flow map. (b) Unblended flow map. (c) Rasterised flow map. Solid cyan circle is Loebensteinstraße. Colour scale is traffic flow, label is traffic flow error. }
\label{fig:flow9}
\end{figure}

If we zoom into Loebensteinstraße (solid cyan circle), then we observe that in Figure~\ref{fig:flow9}(a) the locally aligned map displays the aggregated flows on single linear road segments in comparison to the multiple  road segments of the unblended flow map in Figure~\ref{fig:flow9}(b), and the pixelated  road segments of the rasterised flow map in Figure~\ref{fig:flow9}(c). Whilst we do not have a reference flow map for these empirical trajectories, like we had for the synthetic Los Angeles trajectories, we are still able to compute a proxy flows based on empirical counts of the map matched routes with orthogonal line transects from the flow map road segments. Along Loebensteinstraße, the locally aligned flow map is composed to two road segments (in pink) with 0 and 2 flow errors corresponding to the proxy flows 70 and 58.
For the unblended and rasterised maps, the flow errors vary from 0 to over 100.
Rasterisation some times leads to double counting and thus to larger flow errors, e.g. in the orange region near the intersection, the estimated flow is 365 
whereas the proxy flow is 224, so the flow error is 141. The locally aligned map does not suffer from these rasterisation effects or from an insufficiently simplified topology, so its flow aggregation is more accurate.

For our purposes, the synthetic trajectories in Los Angeles, and the empirical trajectories collected from a single driver in Hannover, can be considered to represent the complete driver population. So the resulting flow maps describe the complete traffic flows. This is not the case for the usual crowd-sourced trajectory data which are drawn from a highly biased sample of the complete driver population, and so the resulting flow map is a highly biased estimate of the complete traffic flow. We leave it to future work to apply the necessary adjustments to compute an unbiased estimate of the complete traffic flow map \citep{iqbal2014,miller2020}.

\subsection{Computation}

We report the execution times for the flow map computations for the synthetic and empirical trajectories. Since it is difficult to reliably reproduce execution times on different systems, we provide our execution times as a guide only. These are obtained on a stand-alone set-up running Ubuntu 24.04 with 16Gb RAM and 4\,\texttimes\,Intel i5 CPU @ 3.10 GHz. 
We deploy a local dockerised Valhalla routing engine API (\url{https://github.com/gis-ops/docker-valhalla}) \citep{valhalla-docker} rather than relying on a web-based service. Since the local map matching API requests can be efficiently parallelised on a stand-alone machine, we are confident that \code{ST\_ROUTE} is suitably optimised.

For $n=100$ synthetic Culver City trajectories, \code{ST\_ROUTE} executes in 18\,s. For $n=1000$ synthetic trajectories, it executes in 158\,s, which is about 10 times longer for a 10-fold increase in $n$. So it is straightforward to infer a precise execution time of \code{ST\_ROUTE} for large sample sizes from a small sample size. The execution time for \code{ST\_OVERLINE} increases as $n$ increases, but not in the same uniform manner as \code{ST\_ROUTE} as the former also depends on the value of $\varepsilon$.
For $n=100$, the execution times for \code{ST\_OVERLINE} are 94\,s ($\varepsilon=1$\,m), 85\,s ($\varepsilon=5$\,m), 99\,s ($\varepsilon=10$\,m); and for $n=1000$, they are 530\,s ($\varepsilon=1$\,m), 358\,s ($\varepsilon=5$\,m), 409\,s ($\varepsilon=10$\,m). So it is difficult to infer precisely the execution times for \code{ST\_OVERLINE} for larger sample sizes and smaller blend tolerances from smaller $n$ and larger $\varepsilon$. 
This is because the line blending (\code{ST\_LINEBLEND}) is only partially parallelised. 

With a blend tolerance $\varepsilon=5$\,m, the total execution time of the workflow for the $n=100$ Culver City trajectories is 103\,s or $\sim$2\,mins, and for $n=1000$ is 243\,s or $\sim$4\,mins. The total execution time for the $n=1183$ Hannover trajectories with $\varepsilon=4$\,m is 1514\,s or $\sim$25\,mins, reflecting the more complex structure of empirical trajectories compared to synthetic trajectories. Thus locally aligned flow maps are currently suitable for offline analyses of small and medium sized data sets. 
Due to time and computing constraints, (a) data sets with a larger number of trajectories (like the full set of 1.5 million synthetic Los Angeles trajectories) and (b) online computations were out of scope. For these to be computationally feasible, we require additional implementation improvements in the local alignment algorithms, which we leave for future development.

Our other future goal is the integration of the local alignment algorithms into an industry standard web GIS. In the mean time, we make available interactive web maps at \url{https://mvstat.net/traj} which allow the user to freely explore our proposed multi-scale traffic flow maps from the individual road to the city-wide level.

\section{Conclusion}

We have introduced an iterative workflow which takes noisy GNSS vehicle trajectories as input and gives an accurate traffic flow map as output. Local alignment of flow segments is the key innovation.
To carry out this local alignment, we defined a spatial relation to set up local reference flow segments, which then allowed us to align other nearby flow segments to this local reference segment. 
We presented solid evidence for the high level of spatial resolution and accuracy for our proposed locally aligned flow map for both synthetic and empirical trajectories. We demonstrated that they effectively reduce the noise of the input trajectories, and accurately aggregate the traffic flows on all road segments, and so they are a valuable addition to the toolkit for multi-scale traffic flow analysis. 

\begin{appendices}
\section{Details of algorithms}

\subsection{Map matched route}

The inputs to \code{ST\_ROUTE} are the trajectory $G$ and the map matching API $M$.
The output is a map matched route $M(G)$.  
In Step~1, we compute the map matched route $M(G)$ from the trajectory $G$ by calling the map matching API $M$. This map matched route $M(G) = \{\bbm_1, \dots, \bbm_{n_M}\}$ is a linestring with $n_M$ edges, with $n_M+1$ points. 

\begin{algorithm}[!ht]
\caption{\code{ST\_ROUTE} -- Map matched route}
\label{alg:st-route}
\begin{algorithmic}[1]
\Statex {\bf Input:} $G$ trajectory, $M$ map matching API
\Statex {\bf Output:} $M(G)$ map matched route
\State Compute map matched route $M(G)$ from empirical trajectory $G$
\end{algorithmic}
\end{algorithm}

\subsection{Snap node clustering of linestrings}

The inputs to \code{ST\_SNAPNODE} are the $n_\mathcal{F}$ flow linestrings $\mathcal{F} = \{(f_1,\bbf_1), \dots, (f_{n_\mathcal{F}}, \bbf_{n_\mathcal{F}}) \}$ and the snap tolerance $\varepsilon_S$. The output are the  snapped linestrings $\mathcal{F}^*$. 
In Step~1, we extract the boundary points of all flow linestrings. In Steps~2--3, we compute a single linkage clustering on all boundary points, and cut the dendrogram at height $\varepsilon_S$, resulting in $C'$ clusters. In Steps~4--7, within each of these single linkage clusters, we compute a complete linkage clustering, and cut the dendrogram at height $\varepsilon_S$. This divides each single linkage cluster into $C''$ clusters, where all cluster members are at most $\varepsilon_S$ distance from each other.
In Steps~8--11, we compute the weighted spatial centroid in each of the $C''$ complete linkage clusters, weighted by the traffic flow values. In Step~12, we renumber the cluster labels from the above nested clusterings to approximate a 1-pass complete linkage clustering into $C^*$ clusters. In Steps~13--16, within each of these $C^*$ clusters, for all points of the corresponding flow linestrings which are closer than $\varepsilon_S$ distance to the centroid, we snap them to the centroid. In Steps~17--18, we collate and sort all the $n_{\mathcal{F}^*}$ snapped linestrings $\mathcal{F}^* = \{(f_1^*,\bbf_1^*), \dots, ( f_{n_{\mathcal{F}^*}}^*, \bbf_{n_{\mathcal{F}^*}}^*)\}$. 

\begin{algorithm}[!ht]
\caption{\code{ST\_SNAPNODE} -- Snap node clustering of linestrings}
\label{alg:st-snapnode}
\begin{algorithmic}[1]
\Statex {\bf Input:} $\mathcal{F}$ flow linestrings, $\varepsilon_S$ snap tolerance  
\Statex {\bf Output:} $\mathcal{F}^*$ snap node clustered flow linestrings
\State Extract boundary points $B(\mathcal{F}) := \{ \Start(\bbf_1), \End(\bbf_1), \dots, \Start(\bbf_{n_\mathcal{F}}), \End(\bbf_{n_\mathcal{F}}) \}$
\State Compute hierarchical clustering with single linkage on $B(\mathcal{F})$
\State Cut single linkage dendrogram at height $\varepsilon_S$ to compute $C'$ cluster labels 
\For{$i := 1$ to $C'$}
\State Extract $i$th cluster of boundary points $B_i' := \{ \bb_1',  \dots, \bb_{n_{B'}}'\}$
\State Compute hierarchical clustering with complete linkage on $B_i'$ 
\State Cut complete linkage dendrogram at height $ \varepsilon_S$ to compute $C''$ cluster labels 
\For{$j := 1$ to $C''$}
\State Extract $j$th cluster of boundary points $B_j'' := \{ \bb_1'', \dots, \bb_{n_{B''}}''\}$ 
\State Extract corresponding flows $\{ \bbf_1'', \dots, \bbf_{n_{B''}}'' \}$ from $\mathcal{F}$
\State Compute $\bb_j^*$ := weighted centroid of $\{ \bbf_1'', \dots, \bbf_{n_{B''}}''\}$, weights := $f_1'', \dots, f_{n_{B''}}''$ 
\EndFor
\EndFor 
\State Renumber collated complete linkage cluster labels to unique $C^*$ labels 
\For{$i := 1$ to $C^*$}
\State Extract corresponding flow linestrings $\mathcal{F}_i^* := \{ (f_1^*,\bbf_1^*), \dots, (f_{n_{B^*}}^*, \bbf_{n_{B^*}}^*)\}$ from $\mathcal{F}$ 
\State Snap points of $\{\bbf_1^*, \dots, \bbf_{n_{B^*}}^*\}$ within $\varepsilon_S$ distance of cluster centroid  to $\bb^*_i$
\State $\mathcal{F}_i^*$ := rejoin snapped and unsnapped points into linestrings 
\EndFor
\State Collate snapped linestrings $\{ \mathcal{F}_1^*, \dots \mathcal{F}_{C'^*}^* \}$ into $\mathcal{F}^* := \{(f_1^*,\bbf_1^*), \dots, (f_{n_{\mathcal{F}^*}}^*, \bbf_{n_{\mathcal{F}^*}}^*)\}$ 
\State Sort $\mathcal{F}^*$ by descending flow and length
\end{algorithmic}
\end{algorithm}

\subsection{Split nodes at interior intersections of linestrings}

The inputs to \code{ST\_SPLITNODE} are the flow linestrings $\mathcal{F}$ and the node splitting type $S$. The output are the split linestrings with added nodes $\mathcal{F}^*$. 
We deploy, for $S=$ ``subdivision'' in Steps~1--3, \code{to\_spatial\_subdivision} in the \pkg{sfnetworks} package \citep{sfnetworks}, and for $S=$ ``unary'' in Steps~4--5, \code{geos\_unary\_union} in the \pkg{geos} package \citep{geos}. 

\begin{algorithm}[!ht]
\caption{\code{ST\_SPLITNODE} -- Split nodes at interior intersections of linestrings}
\label{alg:st-splitnode}
\begin{algorithmic}[1]
\Statex {\bf Input:} $\mathcal{F}$ flow linestrings, $S$ node splitting type  
\Statex {\bf Output:} $\mathcal{F}^*$ node split flow linestrings
\State Initialise local network $\mathcal{N}^*$ with linestrings $\mathcal{F}$
\If {$S==$ ``subdivision''}
\State $\mathcal{F}^* :=$ \code{sfnetworks::to\_spatial\_subdivision}$(\mathcal{N}^*)$ 
\ElsIf {$S==$ ``unary''}
\State $\mathcal{F}^* :=$ \code{geos::geos\_unary\_union}$(\mathcal{N}^*)$ 
\EndIf 
\end{algorithmic}
\end{algorithm}

\subsection{Blend candidate linestrings onto reference linestring}

The inputs to \code{ST\_LINEBLEND} are the reference linestring $\bbf_\rref$, the $n_{\mathcal{F}_\cand}$ candidate linestrings $\mathcal{F}_\cand$, and the blend tolerance $\varepsilon$. The output are the modified reference linestring and $n_{\mathcal{F}_\cand}$ modified candidate flow linestrings, all with added interior points for accurate calculation of exactly equal linestring segments. 
In Step~1, we initialize a local network graph $\mathcal{N}^*$ with the reference linestring. In Steps~2--3, we extract all points of the candidate linestrings, and use the \code{network\_blend} function in the \code{sfnetworks} package (which we denote as \code{ST\_NETWORK\_BLEND}) to blend efficiently these points into the reference linestring \citep{sfnetworks}. 
\code{ST\_NETWORK\_BLEND} projects the candidate linestrings onto the reference linestring, and explicitly adds them to the network, thereby creating new edges in $\mathcal{N}^*$. The result is a local network graph with more, shorter edges and with nodes at the projected candidate points, and whose union is the reference linestring.
In Steps~4--5, we extract the $n_{\mathcal{F}_\cand}$ blended candidate linestrings with these added interior points by applying the shortest path search between the start and end point of each blended candidate linestring along the network graph $\mathcal{N}^*$. 
Step~6 involves collating the blended candidate linestrings. Step~7 is the equivalent for the blended reference linestring. 

\begin{algorithm}[!ht]
\caption{\code{ST\_LINEBLEND} -- Blend candidate linestrings onto reference linestring}
\label{alg:st-lineblend}
\begin{algorithmic}[1]
\Statex {\bf Input:} $\bbf_\rref$ reference, $\mathcal{F}_\cand$ candidate flow linestrings, $\varepsilon$ blend tolerance  
\Statex {\bf Output:} $\bbf^*_\rref$ blended reference, $\mathcal{F}^*_\cand$ blended candidate flow linestrings
\State Initialise local network graph $\mathcal{N}^*$ with $\bbf_\rref$  
\State Extract all points $G_\cand$ of candidate linestrings and flow values $f_\cand$ from $\mathcal{F}_\cand$ 
\State Update network by blending candidate points $\mathcal{N}^* := \code{ST\_NETWORK\_BLEND} (\mathcal{N}^*, G_\cand, \varepsilon)$
\For{$i := 1$ to $n_{\mathcal{F}_\cand}$}
\State  $\bbf^*_{\cand,i}$ :=  shortest path from $\Start(\bbf_{\cand,i})$ to $\End(\bbf_{\cand,i})$  along network $\mathcal{N}^*$
\EndFor
\State Collate $\mathcal{F}^*_\cand := \{ \bbf_{\cand,1}^*, \dots, \bbf_{\cand,n_{\mathcal{F}_\cand}}^* \}$ 
\State $\bbf^*_{\rref}$ :=  shortest path from $\Start(\bbf_{\rref})$ to $\End(\bbf_{\rref})$  along network $\mathcal{N}^*$
\end{algorithmic}
\end{algorithm}

\subsection{Snap candidate-touching linestrings onto reference linestring}

The inputs to \code{ST\_SNAP\_CAND\_TOUCH} are the reference linestring $\bbf_\rref$, the $n_{\mathcal{F}_\cand}$ candidate linestrings $\mathcal{F}_\cand$, the $n_{\mathcal{F}_\candt}$ candidate-touching linestrings $\mathcal{F}_\candt$, and the snap tolerance $\varepsilon_S$. The output are $\mathcal{F}^*_\candt$ the snapped candidate-touching flow linestrings. In Steps~1--8, we iterate over each candidate-touching linestring. In Steps~2--3, we compute the intersections between the candidate-touching linestring and the boundary points of the candidate linestrings, and the respective distances. In Steps~4--7, if this intersection point is within $\varepsilon_S$ distance to the boundary points of  $\bbf_\rref$, then we snap the candidate-touching linestring to the closest boundary point. In Step~8, if the intersection point is not within $\varepsilon_S$ distance, then we snap it to the nearest point on $\bbf_\rref$. 
These snapping operations are similar to those in Steps~15--16 in \code{ST\_SNAPNODE} in Algorithm~\ref{alg:st-snapnode}, and ensure that we maintain connectivity between $\mathcal{F}_\candt$ and $\bbf_\rref$. In Step~9, we collate these snapped linestrings. 

\begin{algorithm}[!ht]
\caption{\code{ST\_SNAP\_CAND\_TOUCH} -- Snap candidate-touching linestrings onto reference linestring}
\label{alg:st-snap-cand-touch}
\begin{algorithmic}[1]
\Statex {\bf Input:} $\bbf_\rref$ reference, $\mathcal{F}_\cand$ candidate, $\mathcal{F}_\candt$ candidate-touching flow linestrings, $\varepsilon_S$ snap tolerance  
\Statex {\bf Output:} $\mathcal{F}^*_\candt$ snapped candidate-touching flow linestrings
\For{$i := 1$ to $n_{\mathcal{F}_\candt}$}
\State $\bg_{\candt,i}^* := \bbf_{\candt,i} \cap \{ \Start(\mathcal{F}_{\cand}), \End(\mathcal{F}_{\cand})\}$ 
\State $d_{i,\Start}^* := \code{ST\_DIST}(\bg_{\candt,i}^*, \Start(\bbf_\rref))$; $d_{i,\End}^* := \code{ST\_DIST}(\bg_{\candt,i}^*, \End(\bbf_\rref))$
\If{($d_{i,\Start}^*\leq \varepsilon_S$ and $d_{i,\Start}^* \leq d_{i,\End}^*$)} 
\State $\bbf^*_{\candt,i} := $ snap $\bbf_{\candt,i}$ to $\Start(\bbf_\rref)$
\ElsIf{($d_{i,\End}^* \leq \varepsilon_S$ and $d_{i,\End}^* \leq d_{i,\Start}^*$)} 
\State $\bbf^*_{\candt,i} := $ snap $\bbf_{\candt,i}$ to $\End(\bbf_\rref)$ 
\Else \ $\bbf^*_{\candt,i} :=$  snap $\bbf_{\candt,i}$ to nearest point of $\bbf_\rref$ 
\EndIf 
\EndFor
\State Collate $\mathcal{F}^*_\candt := \{ \bbf^*_{\candt,1}, \dots, \bbf^*_{\candt, n_{\mathcal{F}_\candt}} \}$ 
\end{algorithmic}
\end{algorithm}

\subsection{Compute line blending priority}

The inputs to \code{ST\_LINEBLEND\_PRIORITY}  are the flow linestrings $\mathcal{F}$, the number of connected edges for the reference linestrings $k$, and the blend tolerance $\varepsilon$. The output is a set of non-overlapping reference linestrings $\mathcal{F}_\rref$, a set of unique candidate linestrings $\mathcal{F}_\cand$, and a set of (potentially repeated) candidate-touching $\mathcal{F}_\candt$ linestrings. 
In Steps~1--2, we set up a network graph from the linestrings $\mathcal{F}$, and then extract $\mathcal{K}$, all simple paths (i.e., all paths composed of connected, unique edges) with $k$-edges. In Steps~3--4, we compute the weighted mean of the $k$ flow values, weighted by the length of individual edges, and assign this to the entire $k$-edge path. The function $\Edges(\cdot)$ extracts the component edges from a $k$-edge path. 
In Step~5, we sort the flow linestrings, in descending order of their flow values and length. 
In Steps~6--8, we construct a maximal set of non-overlapping $k$-edge paths $\mathcal{K}^*$. We initialise $\mathcal{K}^*$ to contain the first path. We iterate through $\mathcal{K}$ and, if this path in $\mathcal{K}$ does not overlap the current $\mathcal{K}^*$, then we add it to $\mathcal{K}^*$. For $k=1$, we bypass Steps~2--8. 
In Steps~9--18, we step through the $\bk_i^* \in \mathcal{K}^*$, starting from the linestring with the highest flow and length. In Steps~11--13, if the current flow linestring $\bk_i^*$ has no incident edges in the candidate linestrings $\mathcal{F}_\cand$, then we set it to be a reference linestring. In Step~13, we search for candidate linestrings for $\bk_i^*$, from those linestrings which are not already reference or candidate linestrings, according to Equation~\eqref{eq:lineblend}. 
In Steps~14--17, if there is at least one candidate linestring, then we update $\mathcal{F}_{\rref}, \mathcal{F}_\cand$, and $\mathcal{F}_\candt$. In Step~18, we extract the candidate $\mathcal{F}_\cand$ and candidate-touching $\mathcal{F}_\candt$ linestring sets. 

\begin{algorithm}[!ht]
\caption{\code{ST\_LINEBLEND\_PRIORITY} -- Compute line blending priority}
\label{alg:st-lineblend-priority}
\begin{algorithmic}[1]
\Statex {\bf Input:} $\mathcal{F}$ flow linestrings, $k$ \#edges, $\varepsilon$ blend tolerance
\Statex {\bf Output:} $(\mathcal{F}_\rref, \mathcal{F}_\cand, \mathcal{F}_\candt)$ reference, candidate, candidate-touching linestrings 
\State Initialise network graph $\mathcal{N}^*$ from linestrings $\mathcal{F}$ 
\State Extract all simple paths of length $k$ edges $\mathcal{K} := \{\bk_1, \dots, \bk_{n_\mathcal{K}} \}$ from  $\mathcal{N}^*$ 
\For{$i := 1$ to $n_\mathcal{K}$} 
\State {$f_i$ :=  weighted mean of $k$ flows, weight := len(\Edges($\bk_i$))}
\EndFor
\State Sort $\mathcal{K}$ by descending flow and combined length

\State Initialise $\mathcal{K}^* := \{ \bk_1 \}$ 
\For{$i := 2$ to $n_{\mathcal{K}}$}
\If{($\bk_i \cap \mathcal{K}^*= \{\}$)} $\mathcal{K}^* := \mathcal{K}^* \cup  \{\bk_i\}$
\EndIf
\EndFor

\State Initialise $\mathcal{Q} := \mathcal{F}_\rref := \mathcal{F}_\cand := \mathcal{F}_\candt :=\{\}$
\For{$i := 1$ to $n_{\mathcal{K}^*}$}
\If{($\Edges(\bk_i^*) \notin \mathcal{F}_\cand $)}
\State Set reference linestring $\bk_{\rref} := \bk_i^*$
\State Search $\mathcal{F}_{\cand}^* := \{ \bbf \in \mathcal{F} \backslash ( \Edges(\mathcal{F}_\rref ) \cup \{ \bk_\rref \} \cup \mathcal{F}_\cand) \colon \bbf \subseteq \code{ST\_BUFFER}(\bk_\rref, \varepsilon)\}$
\If{$\mathcal{F}_\cand^* \neq \{ \}$} 
\State Update $\mathcal{F}_\rref := \mathcal{F}_\rref \cup \{ \bk_\rref \}, \mathcal{F}_\cand := \mathcal{F}_\cand \cup \mathcal{F}_\cand^*$
\State Search $\mathcal{F}_{\candt} := \{ \bbf \in \mathcal{F} \backslash \{ \Edges(\mathcal{F}_\rref) \cup \mathcal{F}_\cand \} \colon \code{ST\_TOUCHES}(\mathcal{F}_{\cand}^*, \bbf) \} $
\State Update $\mathcal{Q} := \mathcal{Q} \cup \{(\{\bk_\rref\}, \mathcal{F}_\cand^*, \mathcal{F}_\candt^*)\}$ 
\EndIf
\EndIf
\EndFor
\State Extract $\mathcal{F}_\cand := \{ \mathcal{F}^*_{\cand, 1}, \dots, \mathcal{F}^*_{\cand, n_\mathcal{Q}}\}, \mathcal{F}_\candt := \{ \mathcal{F}^*_{\candt, 1}, \dots, \mathcal{F}^*_{\candt, n_\mathcal{Q}}\}$ from $\mathcal{Q}$ 
\end{algorithmic}
\end{algorithm}

\subsection{Locally align segment flows using line blending}

The inputs to \code{ST\_OVERLINE\_LINEBLEND} are the flow linestrings $\mathcal{F}$, the split node type $S$, the number of connected edges $k$, the blend tolerance $\varepsilon$, and the snap tolerance $\varepsilon_S$. The output are the modified reference linestring $\bbf^*_\rref$ and projected candidate flow linestrings  $\mathcal{F}^*_\cand$, all with added interior points.
In Step~1, we node split the flow linestrings $\mathcal{F}$. In Step~2, we node snap the flow linestrings. In Step~3, we set up the priority for the line blending, and in Step~4 we store all the linestrings that will not be modified by the line blending in $\mathcal{F}^c$. In Steps~5--8, we iterate over each reference linestring. In Step~6, we update the reference and candidate flow linestrings by blending the candidate linestrings.  
In Step~7, we apply \code{ST\_OVERLINE\_PLANR} within each of the $k$-edges of the updated reference linestrings. In Step~8, we compute the weighted mean flow, weighted by the length of the corresponding sub-segments, and assign it as the flow value to all sub-segments. The result from Steps~6--8 is a modified $k$-edge reference linestring with $k$ aggregated flows. 
In Step~9, we snap the candidate-touching linestrings to the reference linestring to maintain connectivity. In Steps~10--11, we collate the modified linestrings from Steps~5--9, with the unmodified linestrings $\mathcal{F}^c$ from Step~4, and sort them in descending flow and length.   

\begin{algorithm}[!ht]
\caption{\code{ST\_OVERLINE\_LINEBLEND} -- Locally align segment flows using line blending}
\label{alg:st-overline-lineblend}
\begin{algorithmic}[1]
\Statex {\bf Input:} $\mathcal{F}$ flow linestrings, $S$ split node, $k$ \#edges, $\varepsilon$ blend tolerance, $\varepsilon_S$ snap tolerance
\Statex {\bf Output:} $\mathcal{F}^*$ aggregated aligned flow linestrings

\State $\mathcal{F} := \code{ST\_SPLITNODE}(\mathcal{F}, S)$
\State $\mathcal{F} := \code{ST\_SNAPNODE}(\mathcal{F}, \varepsilon_S)$
\State $\{\mathcal{F}_\rref, \mathcal{F}_\cand, \mathcal{F}_\candt\} := \code{ST\_LINEBLEND\_PRIORITY}(\mathcal{F}, k, \varepsilon)$
\State $\mathcal{F}^c := \mathcal{F}\backslash (\Edges(\mathcal{F}_\rref) \cup \mathcal{F}_\cand \cup \mathcal{F}_\candt) $
\For{$i := 1$ to $n_{\mathcal{F}_\rref}$} 
\State $\{\bbf_{\rref,i}^*, \mathcal{F}_{\cand,i}^*\} := \code{ST\_LINEBLEND} (\bbf_{\rref,i}, \mathcal{F}_{\cand,i}, \varepsilon)$
\State $\bbf_{\rref,i}^* := \code{ST\_OVERLINE\_PLANR} (\bbf_{\rref,i}^* \cup \mathcal{F}_{\cand,i}^*)$
\State $f^*_{\rref,i}$ := weighted mean of $\bbf^*_{\rref,i}$, weight $:= \mathrm{len}(\Edges(\bbf_{\rref,i}^*))$
\State $\mathcal{F}_{\candt,i}^* := \code{ST\_SNAP\_CAND\_TOUCH} (\bbf_{\rref,i}^*, \mathcal{F}_{\cand,i}, \mathcal{F}_{\candt,i}, \varepsilon_S)$
\EndFor
\State $\mathcal{F}^* := \mathcal{F}_\rref^* \cup \mathcal{F}_\candt^*  \cup \mathcal{F}^c$
\State Sort $\mathcal{F}^*$ in descending order of flow and length
\end{algorithmic}
\end{algorithm}

\subsection{Compute initial locally aligned flow map}

The inputs to \code{ST\_OVERLINE\_INITIAL} are the map matched routes $\mathcal{M}$, the split node type $S$. The output is an initial aligned flow map. In Step~1, we apply \code{ST\_OVERLINE\_PLANR} to the map matched routes to produce a flow map $\mathcal{F}$. 
In Step~2, we apply \code{ST\_SPLITNODE} to this flow map. In Step~3, we apply the flow aggregation. 
 
\begin{algorithm}
\caption{\code{ST\_OVERLINE\_INITIAL} -- Compute initial locally aligned flow map}
\label{alg:st-overline-initial}
\begin{algorithmic}[1]
\Statex {\bf Input:} $\mathcal{M}$ map matched routes, $S$ split node, $\varepsilon_D$ simplify tolerance
\Statex {\bf Output:} $\mathcal{F}$ initial aligned flow map
\State $\mathcal{F} := \code{ST\_OVERLINE\_PLANR}(\mathcal{M})$ 
\State $\mathcal{F} := \code{ST\_SPLITNODE}(\mathcal{F}, S)$
\State $\mathcal{F} := \code{ST\_OVERLINE\_PLANR}(\mathcal{F})$
\end{algorithmic}
\end{algorithm}

\subsection{Compute iterated locally aligned flow map}

The inputs of \code{ST\_OVERLINE} are the initial flow map $\mathcal{F}$, the split node type $S$, the blend tolerance $\varepsilon$, the snap tolerance $\varepsilon_S$, the number of connected edges $\bk$, and the maximum number of iterations $j_{\max}$.  The output is an iteration of a locally aligned flow map. 
Steps~1--6 is the iteration of \code{ST\_OVERLINE\_LINEBLEND} for the $k$-edge reference linestrings. For each $k$ in $\bk$, we iterate until $j_{\max}$ is reached or two consecutive flow maps are identical. Within each iteration, we search for the $k$-edge reference linestrings, blend the candidate linestrings, and snap the candidate-touching linestrings, and compute the flow aggregation. 
In Steps~7--10 is some housekeeping in \code{ST\_OVERLINE\_PRUNE}, where we remove pseudo nodes, remove leaf edges with low flow values, and replace thin loops with non-looped linestrings.

\begin{algorithm}
\caption{\code{ST\_OVERLINE} -- Compute iterated locally aligned flow map}
\label{alg:st-overline}
\begin{algorithmic}[1]
\Statex {\bf Input:} $\mathcal{F}$ initial flow map, $S$ split node, $\bk$ \#edges, $j_{\max}$ max \#iterations, $\varepsilon$ blend tolerance, $\varepsilon_S$ snap tolerance 
\For{$k$ {\bf in} $\bk$}
\State $\mathcal{F}_\prev := \{ \}$; $j := 0$  
\While{$j< j_{\max}$ and $\mathcal{F}_\prev \neq \mathcal{F}$} 
\State $\mathcal{F}_\prev := \mathcal{F}$;  $j := j+1$ 
\State $\mathcal{F} := \code{ST\_OVERLINE\_LINEBLEND} (\mathcal{F}, k, \varepsilon, \varepsilon_S)$
\State $\mathcal{F} := \code{ST\_OVERLINE\_PLANR} (\mathcal{F})$
\EndWhile 
\EndFor
\State $\mathcal{F}_\prev := \{ \}$; $j := 0$  
\While{$j< j_{\max}$ and $\mathcal{F}_\prev \neq \mathcal{F}$} 
\State $\mathcal{F}_\prev := \mathcal{F}$;  $j := j+1$ 
\State $\mathcal{F} := \code{ST\_OVERLINE\_PRUNE} (\mathcal{F})$ 
\EndWhile 
\end{algorithmic}
\end{algorithm}

\end{appendices}

\section*{Data availability statement}

The data that support the findings of this manuscript are openly available. The Hannover trajectories are hosted in the Leibniz University data depository \url{https://doi.org/10.25835/9bidqxvl}, and the Los Angeles trajectories in the Dryad depository \url{https://doi.org/10.5061/dryad.4j0zpc8gf}. 

\section*{Disclosure statement}

The authors report there are no competing interests to declare.

\bibliographystyle{chicago}

%\bibliography{~/Documents/bibliography/biblio.bib} \end{document}

\end{document}